\documentclass{CUP-JNL-DTM}%

\usepackage{graphicx}
\usepackage{multicol,multirow}
\usepackage{amsmath,amssymb,amsfonts}
\usepackage{mathrsfs}
\usepackage{amsthm}
\usepackage{rotating}
\usepackage{appendix}
\usepackage[authoryear]{natbib}
\setcitestyle{authoryear,open={},close={}} 
\usepackage{subcaption}
\usepackage{wrapfig}

\usepackage{ifpdf}
\usepackage[T1]{fontenc}
\usepackage{newtxtext}
\usepackage{newtxmath}
\usepackage{textcomp}
\usepackage[dvipsnames]{xcolor}
\usepackage{lipsum}
\usepackage[colorlinks,allcolors=NavyBlue]{hyperref}
\usepackage[hang,marginal]{footmisc}
\usepackage{soul}
\usepackage{etoolbox}
\AtBeginEnvironment{quote}{\quotefont}

\newenvironment{bibliografia}{%
 \vspace{4ex}%
 \section*{\refname}
 \addcontentsline{toc}{section}{\refname}
 \normalsize \list{}{%
 \setlength{\itemsep}{0pt}
 \setlength{\itemindent}{-\parindent}
 \setlength{\leftmargin}{\parindent}
 \setlength{\parsep}{\parskip}}
 \vspace{1.5ex}%
 }
 {\endlist}

\theoremstyle{definition}

\numberwithin{equation}{section}

\renewcommand\refname{Bibliography}

\jname{Atti del XLIV Congresso Nazionale Sisfa - Firenze 2024}
\artid{x}
\jyear{2025}

\begin{document}

\begin{Frontmatter}
\jdoi{10.\reflectbox{4\mbox{\st{$7$}}SIS}/0001}
\title[Article Title]{The history of the Arcetri Physics Institute from the 1920s to the end of the 1960s}

\author[1]{Daniele Dominici}

\orcid{0000-0002-4701-2012}

\address[1]{\orgname{University of Florence, Department of Physics and Astronomy} \& \orgname{INFN - Sezione di Firenze}, \orgaddress{\city{Sesto Fiorentino (FI)},\email{ dominici@fi.infn.it}}}

\authormark{D. Dominici}

\abstract{The history of the Arcetri Institute of Physics at the University of Florence is analyzed from the beginning of the 20th century to the 1960s. Thanks to the arrival of Garbasso in 1913, not only did the Institute gain new premises on Arcetri hill, but also hosted brilliant young physicists such as Rita Brunetti, Enrico Fermi, Franco Rasetti in the '20s and Enrico Persico, Bruno Rossi, Gilberto Bernardini, Daria Bocciarelli, Lorenzo Emo Capodilista, Giuseppe Occhialini and Giulio Racah in the '30s, engaged in the emerging fields of Quantum Mechanics and Cosmic Rays. This internationally renowned \textit{Arcetri School} dissolved in the late 1930s mainly for the transfer of its protagonists to chairs in other Italian or foreign universities. After the war, the legacy was taken up by some students of this school who formed research groups in the fields of nuclear physics and elementary particle physics. As far as theoretical physics is concerned, after the Fermi and Persico periods, these studies enjoyed a new expansion in the sixties thanks to the arrival of Raoul Gatto who created in Arcetri the first Italian school of theoretical physics.}

\keywords{Arcetri, Fermi, Garbasso, Rossi, University of Florence}

\end{Frontmatter}

\vspace{2ex}
\section{The Garbasso period}
Aim of this note is to analyze the history of the Arcetri Physics Institute of the University of Florence from 1920s to 1960s; previous studies, which mainly cover the golden period of the '30s, are contained in (\hyperref[man86]{Mandò, 1986}; \hyperref[bon06]{Bonetti \& Mazzoni, 2006}; \hyperref[bon07]{Bonetti \& Mazzoni, 2007}).

In 1859 the provisional government of Tuscany, which took office after the departure of the Lorraine family, created the Istituto di Studi Superiori, Pratici e di Perfezionamento (ISSPP), an institute to initiate young people into post-graduate studies, with four sections, one medical, one scientific, one philological and one philosophical. The chair of Physics was often vacant until the appointment of Antonio Ròiti in 1880. Ròiti, who had graduated in Mathematics from Pisa in 1869 under the direction of the mathematician Enrico Betti, after his call to Florence was mainly interested in electrical measurements, particularly in the study of the unit of measure of the electric resistance. 	

In 1913, after Ròiti's retirement, Antonio Garbasso was appointed as the chair of Experimental Physics. After graduating from Turin in 1892, he did research experience in Germany with Hertz and Helmholtz and held teaching positions in Turin and Pisa; in 1903 he was called to the chair of Experimental Physics in Genoa, where he remained until he moved to Florence. He worked on optics, and spectroscopy and explained the Stark Lo Surdo effect within the Bohr theory. 

Two years after he arrived in Florence, Garbasso succeeded in obtaining funding from the town council to move the Physics Laboratory from the premises in the center of Florence to a new building on Arcetri hill which already was housing the Astronomical Observatory, not far from Villa Il Gioiello, where Galileo had lived from 1631 until he died in 1642. Garbasso was also very interested in technological applications, as shown by his initiative to open in 1918 the Laboratory of Practical Optics and Precision Mechanics which later became the premises of the National Institute of Optics.

Garbasso was also an important public figure: he was Mayor of Florence from 1920 to 1927, standing with the National Bloc, made up of liberals, radicals, republicans, and reformists. In the field of physics, he was President of the SIF (Italian Physical Society) for two periods (1912-1914, 1921-1925) as well as of the Committee of Astronomy, Mathematics and Physics of the CNR (National Research Council). In addition to the construction of the new building, Garbasso started by recruiting brilliant young physicists, in 1922, as assistant, Franco Rasetti, who had just graduated from Pisa with a thesis on spectroscopy with Luigi Puccianti, and in 1924 Enrico Fermi, as lecturer to cover courses in Theoretical Mechanics and Mathematical Physics.

At that time, Antonio Garbasso's assistants were Rita Brunetti, who was working on spectroscopy in the visible and the X range and was the first woman to win a chair in Physics in Italy in 1926, and Vasco Ronchi, who became Director of the National Institute of Optics when it was founded in 1930.

The course in Theoretical Mechanics was taken by students in the undergraduate courses of Physics, Physics and Mathematics, of Mathematics, and of the two-year engineering preparatory course. Two of the students for engineering studies in the academic year 1925-26 reorganized the notes of the Theoretical Mechanics course into handouts that were printed in 1926\footnote{A book containing these lessons was recently published by Firenze University Press (\hyperref[cas19]{Casalbuoni, Dominici \& Pelosi, 2019})}.

The course in Mathematical Physics, taught in the fourth year of Physics, Physics and Mathematics and of Mathematics, included for 1924/25 the traditional topics of Electrodynamics supplemented by a mention of the new Theory of Relativity. In the following year, the title of the course was changed to Theoretical Physics and Fermi covered notions of Probability, Thermodynamics, and Statistical Mechanics. This course in Theoretical Physics in the academic year 1925/26, together with a similar course held in Naples, were the first with this title in Italy\footnote{For a reconstruction of the institution of the first chairs of Theoretical Physics in Italy, see (\hyperref[lar20]{La Rana \& Rossi, 2020})}.

Rasetti's recollections are very important to reconstruct this period:
\begin{quote}
 My first job was in Florence, and I worked on atomic spectroscopy$\ldots$ The equipment was pretty good for those times - especially for spectroscopy, which was my field. They had a very good spectrograph and spectroscope; we had an excellent Rowland grating in the Rowland mounting. And I didn't have much teaching to do, because [Antonio] Garbasso gave the physics course (\hyperref[goo01]{Goodstein, 2001}).
 \end{quote}
 On Garbasso, Rasetti says:
 \begin{quote} 
  Garbasso had been a good physicist, but when I knew him he was only interested in politics. He was the mayor of Florence$\ldots$ He gave his course in elementary physics and he was quite intelligent at it. And later Fermi explained to him what we were doing and he understood because he was intelligent. I mean, he knew the classical theory - he didn't know much about the quantum theory, because that had come after he lost direct interest in physics. But he followed what we were doing, and he was a very pleasant person (\hyperref[goo01]{Goodstein, 2001}).
\end{quote}

Enrico Fermi's time in Florence was short but very fruitful: he wrote several articles with Rasetti, in particular a paper on the effect of weak but high-frequency magnetic fields on the depolarization of resonance light in mercury vapor. In 1926 Fermi published the work on the Fermi-Dirac statistics, which made him internationally famous. Since 1923 he has been interested in Statistical Mechanics and particularly in the problem of the absolute constant of the entropy of a perfect gas (\hyperref[cor00]{Cordella \& Sebastiani, 2000}; \hyperref[cas23]{Casalbuoni, 2023}). The new factor, that enabled him to discover the Fermi-Dirac statistics, was the Exclusion Principle formulated by Pauli in 1925. Fermi's great merit was having applied Pauli's Principle, which until then had been advanced for the interpretation of spectroscopic phenomena, to a general system of particles. Paul Dirac also arrived at the same result in August 1926, by relating the two quantum statistics to the two possibilities of the wavefunctions of a system being symmetric or antisymmetric under the exchange of the coordinates of two identical particles. After the publication of Fermi's work, important applications by Thomas (1926), independently of Fermi (1927), treating the internal electrons of a heavy atom with statistics, and by Fowler (1926), on white dwarfs and Sommerfeld (1928) on conduction in metals followed.

In 1926 Corbino succeeded in having the first chair of Theoretical Physics and, after the selection process, Fermi was appointed to Rome, Enrico Persico to Florence and Aldo Pontremoli to Milan. The arrival of Persico in Florence on the chair of Theoretical Physics was of great importance for his extraordinary teaching skills and his contribution to the spread of Quantum Mechanics. 

After Fermi also Rasetti left for Rome, then Garbasso hired Bruno Rossi and Gilberto Bernardini in 1928, and thanks to the graduations of Giuseppe "Beppo" Occhialini in 1929 (advisor Rossi), Giulio Racah (advisor Persico) and Daria Bocciarelli (advisor Rossi) in 1931 and Lorenzo Emo Capodilista in 1932, a group of young people who greatly boosted cosmic ray research was formed\footnote{For Bruno Rossi's contribution to the physics of cosmic rays, see (\hyperref[bon11]{Bonolis, 2011}) and talk at this conference.}.

In Arcetri Rossi developed the famous coincidence circuit, consisting of triodes allowing the detection of triple or higher coincidences of ionizing particles. In this device, the current flow in the circuit is interrupted only when all counters are working at the same time. With this new electronic coincidence apparatus at his disposal, improving sizably the time resolution of Bothe's and Kohlhörster's experiment, Rossi experimented with magnetic lenses to measure the charge of the corpuscles that made up the cosmic rays. In the summer of 1930, when Rossi was in Berlin with Walter Bothe, he conjectured the existence of an east-west asymmetry in the distribution of cosmic rays, due to the effect of the earth's magnetic field, which predicted the arrival of more particles from the east or west of the magnetic meridian depending on whether the particle's charge was negative or positive. However, his experiment on this effect performed in Florence proved negative within the errors. 

A particularly important consequence of the cosmic ray studies stemmed from the lack of expertise on cloud chambers in Italy, which were fundamental instruments for determining the characteristics of particles. Rossi, who had met in the 1930 summer in Berlin Patrick Blackett, Europe's leading expert on the subject, decided to send Occhialini to work there. Occhialini left in 1931, taking the skills acquired in the field of coincidences at Arcetri with the idea of combining the Rossi circuit with the cloud chamber. Blackett and Occhialini obtained their first results in 1933, the most exciting being the discovery of the showers produced by cosmic rays and the formation of electron-positron pairs. Moreover, thanks to the cloud chamber immersed in a magnetic field, it was possible to observe the components of the shower and also determine the sign of the particle charge. Occhialini's contribution was extremely important for the identification of the positron thanks to Rossi's circuit. This enabled him and Blackett to confirm Carl Anderson's discovery of the positron (1932), published just a few months earlier, and the positive and negative electron showers\footnote{For the influence of the Arcetri school on the formation of Giuseppe Occhialini see Tucci contribution at this conference.}.

Racah was not very involved in the experimental research on cosmic rays however he did relevant calculations of bremsstrahlung cross sections from high-energy electrons, of the production of electron-positron pairs, and of hyperfine structures in atoms. From 1932 to 1937, he taught the course Theoretical Physics and then moved to Pisa, and in 1939, because of the racial laws, to Palestine. He often travelled to Rome to collaborate and discuss with Enrico Fermi, Ettore Majorana, and Gian Carlo Wick and had also established an important working relationship with Wolfgang Pauli in Zurich.

\begin{figure}[h]
\vspace{-0.2cm}
\centering
\includegraphics[width=0.99\textwidth]{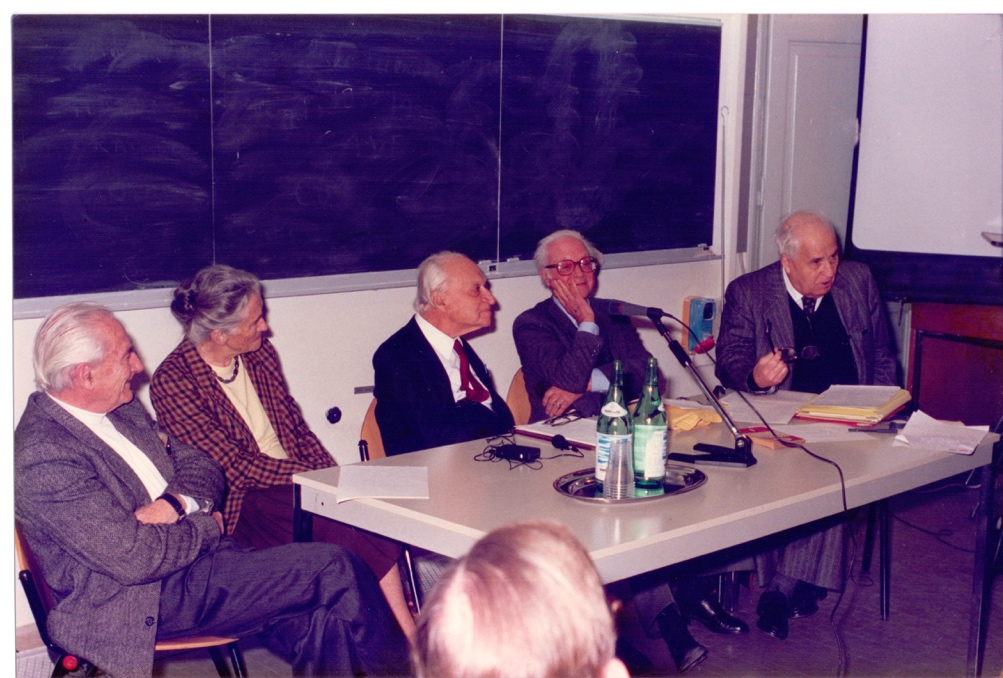}
\caption{Round table at the conference held in Arcetri in 1987 for the 80th birthday of G. Occhialini. From left: Bernardini, Bocciarelli, Rossi, Amaldi, Mandò (courtesy of Pier Andrea Mandò).}
\label{fig:1}
\vspace{-0.3cm}
\end{figure}

It is also important to note that the scientific relations between the Florence and Rome groups in those years were very intense, as remembered by Edoardo Amaldi in a round table at the conference held in Arcetri in 1987 for the 80th birthday of G. Occhialini (\hyperref[bon07]{Bonetti \& Mazzoni, 2007}), (Fig. \ref{fig:1}). Rasetti, who had been Garbasso's assistant in 1922, became Orso Mario Corbino's assistant in 1926, filling the position vacated by Persico, who had been called to Florence to the chair of Theoretical Physics. This exchange of personnel strengthened relations and collaborations between the two groups, which soon achieved international fame working on different topics, cosmic rays in Florence and atomic spectroscopy, the Raman effect and, from 1932, nuclear physics in Rome.

During one of these visits on a weekend in 1933, Bruno Rossi, after a discussion with his Roman colleagues on the effect of the earth's magnetic field, wrote an article on this subject in collaboration with Fermi where they showed that the explanation for the negative result for the east-west effect required large absorption by the atmosphere and that the effect would have been visible near the equator. Rossi, also helped by Garbasso, began to organize a mission to Eritrea to prove this statement. The mission could only be held in 1933 after Rossi moved to Padua. Sergio De Benedetti and Ivo Ranzi also participated in the mission. The experiment definitively confirmed the corpuscular theory of cosmic rays that the prevailing direction of the particles was from the west of the magnetic meridian and that the particles were therefore positively charged. Unfortunately, due to the delay in the organization, the Rossi experiment could take place only after the Alvarez Compton and Johnson ones.

Thanks to the good relationship between the physics institutes in Florence and Rome, when Fermi needed Geiger counters to study neutron-induced radioactivity, the Florentines supplied all their expertise. In particular, the Bothe secret on the fact that his Geiger counters had not a steel but an aluminum wire, which Bothe confessed to Rossi in Berlin in 1930, was from Rossi transmitted to his friends in Florence and Rome (\hyperref[ros87]{Rossi, 1987}). Amaldi also recalls that when Rossi had already moved to Padua, “one weekend in April or May 1934, Bernardini, Occhialini, Daria Bocciarelli, and Emo Capodilista came to Rome and brought us boxes full of Geiger counters and proportional counters: they were a gift to help us in our work... They were beautiful and worked very well, but unfortunately, the geometry was wrong” (\hyperref[bon07]{Bonetti \& Mazzoni, 2007}). An interesting hypothesis has been put forward to this end, that among the counters tested by Fermi and mentioned in the Roman scientist’s notebook, found by Francesco Guerra and Nadia Robotti in Avellino, there were those brought by the Florentines (\hyperref[gue15]{Guerra \& Robotti, 2015}). 

During the 1987 Occhialini conference, it was also remembered by Occhialini and Bernardini the fundamental contribution of Giorgio Abetti, who, as Director of the Observatory, launched the Mathematical, Physical and Astrophysical Seminar of Arcetri. This was a series of conferences by international speakers aimed at all students and professors of the university, which was officially approved by the Faculty in 1932 and ran until 1943. This seminar was of great importance for the young people of the Institute of Physics, because it allowed them to get to know many world-famous scientists, who were invited by Abetti regularly.

\begin{figure}[h]
\centering
\includegraphics[width=0.99\linewidth]{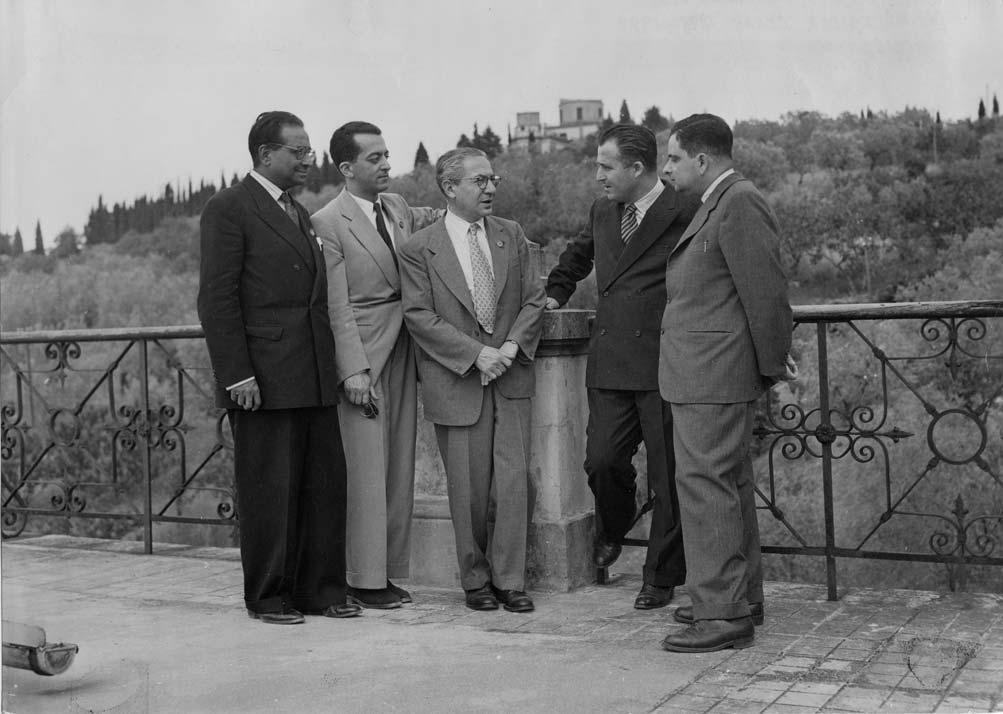} 
\caption{From the right Franchetti, Mandò, Isidor Rabi and two other visitors on the Arcetri terrace (University of Florence, Department of Physics and Astronomy, Scientific and Technological Hub).}
\label{fig:2}
\vspace{-0.3cm}
\end{figure}

\begin{figure}[hb]
\centering
\includegraphics[width=0.99\linewidth]{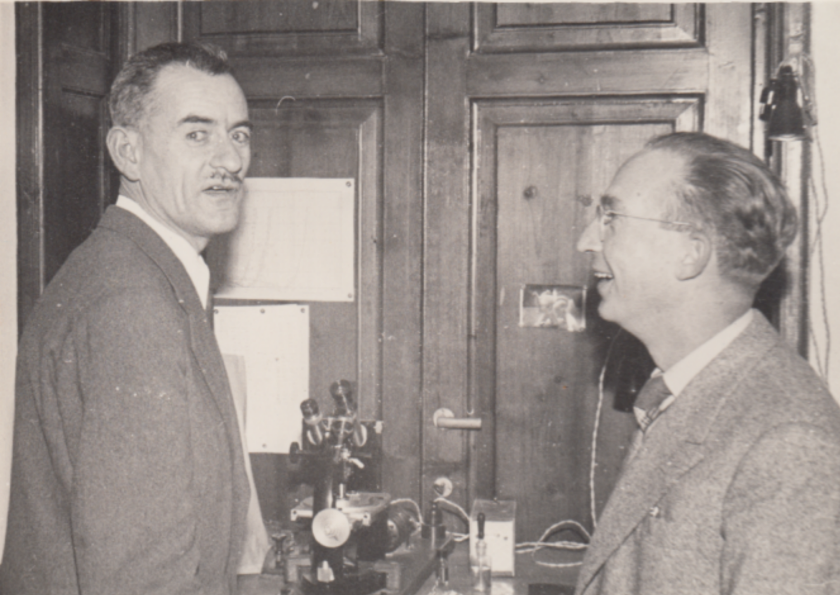}
\caption{Della Corte (right) with Louis Leprince-Ringuet in Florence in 1954 ( University of Florence, Historical Archives of the Sciences Library, \textit{Michele Della Corte})}
\label{fig:3}
\end{figure}

The commencement of nuclear research in Italy dates back to 1933 and the institutes involved included Rome, Padua, and Florence. It has to be stressed that this program, supported by the National Research Council, assigned the Florentine institute the task of studying the excitation of neutrons in various elements with $\alpha$ particles of different energies, as well as the disintegrations produced by neutrons as they pass through matter (\hyperref[gue15]{Guerra \& Robotti, 2015}). This program involved Bernardini, Bocciarelli and Capodilista, but was essentially carried out abroad by Bernardini and Emo Capodilista in Berlin Dahlem, where Lise Meitner was also working. The group in Padua was supposed to focus on cosmic rays while the group in Rome was supposed to focus on $\gamma$-spectroscopy. The Rome group, as is well known, under the leadership of Fermi was to discover neutron-induced radioactivity in 1934, paving the way for nuclear fission.

In 1932, Rossi moved to the chair of Experimental Physics at the University of Padua. After his departure, the research activity on cosmic rays in Arcetri was carried on by Bernardini, Bocciarelli, Capodilista and Franchetti. The activity of these researchers was also devoted to the study of induced radioactivity, like for example in the polonium-beryllium system (polonium emits $\alpha$ particles that hit beryllium that becomes neutron emitter). Unfortunately, at the end of the 1930s, this golden age of Florentine physics came to an end. In 1937, Bernardini left Florence for the chair of the University of Camerino; Occhialini, after his return in 1934, 1937 left for Brazil to escape the fascist regime sizing the opportunity of Gleb Wataghin's invitation and contributing to the birth of the Brazilian school of particle physics in Säo Paulo. He returned to Europe in 1944 and collaborated with the Brazilian Cesare Lattes and with Cecil Powell on the discovery of the pion in 1947. Daria Bocciarelli, in 1938, moved to the Istituto Superiore di Sanità in Rome, where she played a very important role. Racah moved to Pisa in 1937, after winning the second Italian selection for professorships in Theoretical Physics. Emo Capodilista abandoned his scientific career, after a period in the United States at the Lawrence Radiation Laboratory. 

Added to all this in 1933 Antonio Garbasso died and was replaced by Laureto Tieri, an old-style professor who was not much interested into the new physics. The atmosphere at the Institute had changed, \textit{the spirit of Arcetri} was no longer there. Perhaps these young researchers would have gone anyway, but certainly, the change of the human environment at the Institute was largely responsible for it. By the end of the 1930s, Manlio Mandò, Michele Della Corte, and Giuliano Toraldo di Francia had graduated from Arcetri, and, together with Simone Franchetti and Nello Carrara, they were to contribute to the rebirth of Florentine physics in the post-war period (Figs. \ref{fig:2} and \ref{fig:3}).

\section{The Second World War period}

In the academic year 1937/38, Franchetti became Assistant Professor and lecturer in Theoretical Physics in Florence, where he began his studies on the physics of the nucleus, focusing particularly on the interaction of $\gamma$-rays with the matter, and later on the spectra of $\mu$ leptons or, as they were then called, mesons. In 1937, Tito Franzini arrived from the University of Pavia and Vincenzo Ricca from the University of Messina. Tito Franzini moved as Assistant Professor to Florence to take over the position vacated by Mandò who, in the meantime, had become Assistant Professor in Palermo, where Emilio Segré had been called as Professor from 1935. After the departures described at the end of the previous Section, Simone Franchetti, the last to remain at the Institute from the time of Bernardini, was expelled from the Florentine University due to the racial laws, because, although his mother was Catholic, his father was Jewish. 

In addition to the Director Laureto Tieri, several young graduates remained at the Institute of Physics, including Carlo Ballario and Della Corte, who, together with Tito Prosperi in 1940, experimented on the absorption of cosmic rays under rock in the Tuscan-Emilian Apennines. The experiment took place in a shaft of the tunnel of the Florence-Bologna railway, which descended a slope to a depth of 200 metres to the underground station called Precedenze. The experiment aimed to detect the particles that make up cosmic rays, after passing through the layers of matter of the mountain, using Geiger Müller counters and an electrical coincidence circuit. The apparatus was mounted on the carriage of the well and this allowed measurements to be taken at various depths.

Della Corte then left for military service and in 1942, at the Cascine Air War School, he met Air Force Captain Italo Piccagli, a staunch anti-fascist. Piccagli was also an expert atmosphere scientist and had presented his studies at the Mathematical, Physical, and Astrophysical Seminar. In the period immediately preceding 8 September 1943, Piccagli proposed to Della Corte the opportunity to transfer the instruments and apparatuses of the Meteorology and Air Navigation Laboratories to Arcetri to save them from requisition by the German army. The equipment was hidden in the Institute of Physics and Arcetri Observatory. The Institute was then searched by the German SS, who had been alerted to the presence of air force material on the premises. Fortunately, the material had been well hidden and the Germans only took away a few items. A few months later, Della Corte and Ballario, again following an invitation by Captain Piccagli, joined Radio CORA, a clandestine radio station promoted by the Radio Commission of the Action Party. The radio, which had the task of transmitting relevant information to the allied commands and partisan troops, had several bases in the city, including the Institute of Physics in Arcetri. The experience of Radio CORA tragically ended on 7 June 1944; the Nazis located the radio on the premises in Piazza d'Azeglio in Florence and stormed in. Luigi Morandi, who was on duty at the radio transmitter managed to fight back and kill a German but was wounded and died a few days later. Despite being tortured, Piccagli and the lawyer Enrico Bocci took the entire responsibility for the organization, exonerating the others. Piccagli was killed by German soldiers in the hills above Florence on 12 June 1944, together with other anti-fascists. Enrico Bocci was probably killed by the SS, but his body was never found. Della Corte and Ballario were saved by the sacrifice of Bocci and Piccagli and because they were not on radio duty that day. From 1944 to 1947, Ballario was Assistant Professor at the University of Bologna before moving to Rome, where he collaborated with Amaldi and Bernardini's group. Della Corte continued his academic career in Florence. 

Simone Franchetti's expulsion from the university ended on 28 December 1944 when he resumed service at Arcetri. Mandò, called to arms and sent to Libya in 1939, in December 1940 was badly wounded in battle and captured by the British army. He was operated on and treated by the British in the military hospital in Alexandria and then transferred to prison camps in India. He finally returned to Italy on 1st July 1946.

\section{After the war until the ‘60s}

In this section the birth of three areas of physics research in Florence in the 1960s, high energy physics, physics of the nucleus and finally theoretical physics is reviewed. For the microwave and laser physics activities, that were born and developed in Florence around the figures of Nello Carrara and Giuliano Toraldo di Francia, see (\hyperref[cas22]{Casalbuoni, Dominici \& Mazzoni, 2022}).

In 1951, on the initiative of Edoardo Amaldi and Gilberto Bernardini and with the support of Gustavo Colonnetti, President of the CNR, a new important research agency, the Istituto Nazionale di Fisica Nucleare (National Institute of Nuclear Physics), was born to coordinate theoretical and experimental research activities in nuclear physics and cosmic rays. Its first four Sections were those of Rome, Padua, Milan, and Turin. In Florence, a Subsection was created in 1952, and was directed by Franchetti until 1966, by Renato Angelo Ricci in 1967/68, and then by Mandò until 1972, when it was transformed into a Section.

\subsection{Experimental physics}

The Arcetri school's research into cosmic rays, initiated by Rossi, was continued by Della Corte, who, in 1950, with the support of CNR and Della Riccia Foundation grants, went to Paris to work with Louis Leprince-Ringuet's group at the École Polytechnique, to learn the nuclear plate technique. Upon his return, he created the Florentine \textit{plate group}, a team performing particle physics experiments with nuclear emulsions. The group's first research was on trace formation and the determination of the charge and the mass of the particles (\hyperref[car14]{Cartacci, 2014}).

The Florence group, joined later by Anna Maria Cartacci and Pier Giorgio Bizzeti, then switched from research on cosmic rays to research with accelerators, by collaborating with groups from other Italian universities at CERN, the new European facility for particle and nuclear physics. After the study of positive pion-proton scattering at 8.3 MeV to determine phase shifts, the Florence-Genoa-Turin collaboration studied 25 GeV proton interactions at CERN's Proton Synchrotron, determining the characteristics of these interactions; the Parma-Florence collaboration on the other hand studied the final products in the interactions of strange mesons in nuclear emulsions. In 1964, the group abandoned the technique of nuclear emulsions and switched to the analysis of bubble chamber frames in experiments at CERN. 

At the end of the 1960s, Della Corte, having become increasingly interested in the application of physics to medicine, switched to nuclear medicine, abandoning elementary particle research; however, Giuliano Di Caporiacco and Giuliano Parrini joined the group. In collaboration with the groups in Bologna, Bari, and the Institut de Physique Nucléaire in Orsay, the Florentine group published results on the research of new particles in resonant pion systems, produced in interactions of the 5.1 GeV positive pion beam of CERN's Proton Synchrotron in a bubble chamber filled with deuterium. A new collaboration with the groups of Milan, Bologna, and Oxford continued studying the production of new states in interactions of $\pi^-$ of 11.2 GeV in a bubble chamber filled with hydrogen. The group would later return to the emulsion technique in the search for particles containing a charm-type quark, carried out at CERN in the 1970s. 

Before analyzing the history of the nuclear group, it is worthwhile to mention also the figure of Tito Fazzini, who graduated with Franchetti in 1947 on the energy spectrum of $\mu$, in 1957 moved to CERN, where, three years after CERN birth, the 600 MeV CERN synchrocyclotron (SC) started accelerating protons. In August 1958, Fazzini, Giuseppe Fidecaro, Alec Merrison, Helmut Paul and Alvin Tollestrup proved the rare pion decay in electron neutrino by measuring at the SC the ratio R=$\frac{\Gamma(\pi^+ \rightarrow e^+\nu_e)}{\Gamma(\pi^+ \rightarrow \mu^+\nu_\mu)}>5x10^{-5}$, a value consistent with the V-A expectation of weak interaction theory. Fazzini returned to Florence in 1962, where he focused on research into nuclear physics with Mandò’s group.

At the end of the 1950s, Manlio Mandò, after moving from Bologna to Florence as an Assistant Professor of Experimental Physics and having carried out research on cosmic rays in the immediate post-war period, succeeded in setting up an experimental group to work on nuclear physics. In the years that followed, this group became of international importance thanks partly to its fruitful contacts with centers of excellence for nuclear physics in the United States, Japan and especially Germany. In addition to Manlio Mandò, Tito Fazzini, Piergiorgio Bizzeti (after the initial period with the plate group), Anna Maria Bizzeti Sona, Mario Bocciolini, Giuliano Di Caporiacco (who later joined the plate group) and, from 1962-63, the new graduates Pietro Sona, Paolo Maurenzig, Nello Taccetti and Paolo Blasi were part of the nuclear physics group and in the early 1960s turned to experimental nuclear physics with accelerators, using a PN400 Van de Graaff accelerator, which provided a terminal voltage of 400 kV and could accelerate protons and deuterons to be sent to a target, to produce neutrons and $\gamma$-rays (\hyperref[tac17]{Taccetti, 2017}).

The main activities carried out were the production of isomeric states and the measurement of photo-production cross-section fluctuations. The group also worked on the development of new detectors. When Renato Ricci moved to the University of Florence in 1965, he founded the aforementioned group of young experimental physicists, which was studying the use of new Silicon-based detectors to detect electrons and $\alpha$ particles, and new Germanium-based detectors to detect $\gamma$-rays. These detectors made it possible to achieve much better energy resolutions than could be obtained with thallium-doped sodium iodide detectors. At the end of 1966, Ricci proposed to the group the participation in nuclear spectroscopy research at the Legnaro Laboratory (Padua), which was to become the INFN's Legnaro National Laboratories dedicated to nuclear physics in 1968. This marked the start of the Florentine nuclear group's long collaboration with the Legnaro Laboratory.

The PN400 accelerator was decommissioned at the end of the 1960s. The Florentine group then managed to obtain from INFN the KS3000 electron accelerator, a 3 MV Van de Graaf, which had been the injector of the Frascati Electron Synchrotron, decommissioned in 1968-69. The accelerator, which arrived in Florence in 1971, was transformed into a positive ion accelerator with the important contributions of Tito Fazzini, Giacomo Poggi and Nello Taccetti and the group technicians. The new accelerator, renamed KN3000, was used in the 1970s to perform nuclear spectroscopy measurements and parity violation experiments in nuclei. Its career came to a close with participation, under the leadership of Pier Andrea Mandò, in the initial phase of the programme of nuclear physics applied to the environment and cultural heritage. This initial activity gave birth to the group that now works at the Laboratory of Nuclear Techniques Applied to Cultural Heritage (LABEC) at the Scientific Hub of the University of Florence in Sesto Fiorentino.

\subsection{Theoretical physics}

To conclude, I am going to review the activities of the theoretical physics group. After his initial studies of nuclear physics and cosmic rays, from the 1950s onwards, Franchetti concentrated on the theoretical analysis of condensed states, especially liquids, and was among the pioneers in studying liquid helium.

After Persico's departure in 1930, the chair of Theoretical Physics remained vacant until 1958, when Giacomo Morpurgo was appointed. When in 1962 Morpurgo moved to Genoa the chair was assigned to Raoul Gatto - on the figure of Raoul Gatto see (\hyperref[cas18]{Casalbuoni \& Dominici, 2018}; \hyperref[bat19]{Battimelli, Buccella \& Napolitano, 2019}).

Raoul Gatto graduated at the Scuola Normale Superiore in Pisa in 1951 with a thesis on nuclear shell models, prepared under the guidance of Bruno Ferretti, who held the chair of Theoretical Physics in Rome at the time, and Marcello Conversi, who taught Experimental Physics in Pisa and was his supervisor. In the same year, he obtained a diploma with honors from the Scuola Normale Superiore with a thesis on statistical theories of nuclei, supervised by Tullio Derenzini. After his thesis, Gatto moved to Rome and became assistant of Bruno Ferretti, studying weak hadron decays and associated angular distributions. In 1956 Gatto obtained the "Libera docenza" in Theoretical Physics and moved to the Radiation Laboratory in Berkeley, California, where he remained until 1957. At Berkeley, where Louis Alvarez's group was discovering numerous new particles using a bubble chamber, Gatto continued his work on the phenomenology of hyperons, the symmetries of weak interactions, and the consequences of the violation of parity in weak interactions. In the early 1960s, Gatto became a Full Professor in Cagliari but also spent part of his time at the Frascati Laboratories. As one of the leading theoretical experts in Quantum Electrodynamics, he participated in the research for the first particle collider, or Storage Ring (AdA), the well-known electron-positron machine devised by Bruno Touschek, and in the research for the subsequent ADONE (\hyperref[bon23]{Bonolis, Buccella \& Pancheri, 2023}).

\begin{wrapfigure}{l}{0.6\textwidth}
\vspace{-0.3cm}
\centering
\includegraphics[width=0.96\linewidth]{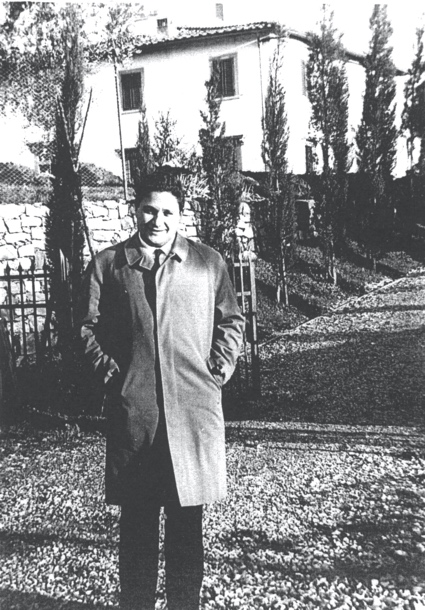} 
\caption{Raoul Gatto in Florence in the 1960s (courtesy of the family).}
\label{fig:4}
\vspace{-0.5cm}
\end{wrapfigure}

In Florence, Gatto (see Fig. \ref{fig:4}) decided to create a group of young theorists, organized in the style of the famous Landau school. He found in Arcetri Marco Ademollo, Emilio Borchi, and Giorgio Longhi, and brought to Florence Guido Altarelli, Franco Buccella, and Giuliano Preparata who had done their theses at the Rome University under his supervision and Enrico Celeghini from Cagliari. Luciano Maiani, who graduated from Rome in 1964 with an experimental thesis on semiconductor detectors, supervised by Giorgio Cortellessa of the Istituto Superiore di Sanità, obtained, thanks to Mario Ageno, a scholarship to work not only in the Istituto Superiore di Sanità but also in Florence with Gatto. The only Florentine student who graduated with Gatto in this early period was Gabriele Veneziano, who, in 1967, began his collaboration with Ademollo on sum rules and dual models, which, in 1968, led him to the formulation of the internationally acclaimed Veneziano model.

The research activity of the group was devoted to phenomenological aspects related to the discovery of new particles like the classification of baryonic and mesonic resonances by using the group SU(3) and its generalization SU(6), determination of their spins from various angular correlations, electron-positron radiative corrections but also to more formal aspects, like dispersion relations, Schwinger terms and non-renormalization theorem. In 1966, Gatto decided to expand the group by recruiting the students who seemed most promising.
On the whole, Gatto's students were truly exceptional, as is shown by the success of their scientific careers, but certainly, the atmosphere and his teaching contributed in no small measure to what has been considered the only true school of theoretical physics ever to have existed in Italy.
Many scientifically important results were obtained, probably the best known of which is the Ademollo-Gatto theorem on certain properties of weak interactions (non-renormalization for the strangeness-violating vector currents to first order in the symmetry-breaking interactions). In the academic year 1967-68, Gatto went on leave to Geneva, then in 1968-69 moved to Padua and again to Rome. In the 1970s, he was called to the University of Geneva where he taught until 1995. At the end of the 1960s, after Gatto's departure, Ademollo, Altarelli, Buccella, Gallavotti, and Preparata left Arcetri for the United States and CERN, Maiani got a permanent position at the Istituto Superiore di Sanit\`a; only Longhi and Celeghini stayed in Florence with a group of young people who had just graduated or were about to (Roberto Casalbuoni, Luca Lusanna, Emanuele Sorace). Let us conclude by quoting (\hyperref[mai17]{Maiani \& Bonolis, 2017}): 

\begin{quote}

 With the end of 1966, the Florence experience came to an end. Although Gatto had done very well in Florence and had become really famous for having grown all these pupils, he was thinking to move... The perfect atmosphere of ’65/’66 was not there anymore... What we achieve in Florence? We certainly participated in the struggle of theoretical physics of those years towards a theory of strong interactions and we had been recognized as useful interlocutors.
\end{quote}

\section{Conclusions}

The history of modern physics at the Arcetri Institute of Physics of the University of Florence traced back to the beginning of the 20th century with the arrival of Antonio Garbasso. Thanks to Garbasso, brilliant groups of young physicists (Rita Brunetti, Enrico Fermi, Franco Rasetti, Enrico Persico, Bruno Rossi, Gilberto Bernardini, Daria Bocciarelli, Lorenzo Emo Capodilista, Giuseppe Occhialini and Giulio Racah) started their scientific carriers in Arcetri. Very important was also the role of Giorgio Abetti, Director of the Arcetri Observatory, who created the Mathematical, Physical, and Astrophysical Seminar, an institution to promote in Arcetri conferences by foreign and Italian physicists and mathematicians.
This internationally renowned Arcetri School ended up in the late 1930s mainly for the transfer of its protagonists to chairs in other universities, but also, to some extent, for the environment created by the fascist regime. After the war, the legacy was taken up by some students of this school who formed research groups in the field of nuclear physics, elementary particle physics, and matter physics. As far as theoretical physics was concerned, after the Fermi and Persico periods, these studies enjoyed a new expansion in the '60s, with the arrival of Raoul Gatto, who created the first real Italian school of theoretical physics.
\section*{Acknowledgments}
The author would like to thank Roberto Casalbuoni and Massimo Mazzoni for their fruitful collaboration on this topic and Luisa Bonolis, Pier Andrea Mandò, and Paolo Rossi for their useful and stimulating suggestions.
The talk is based on (\hyperref[cas21]{Casalbuoni, Dominici \& Mazzoni, 2021}) and on (\hyperref[cas22]{Casalbuoni, Dominici \& Mazzoni, 2022}). A complete list of references can be found in these two works.

\begin{Backmatter}
\vspace{-0.5cm}
\begin{bibliografia}

\label{bat19} \item Battimelli, G., Buccella, F. \& Napolitano, P. (2019). “Raoul Gatto, a great Italian scientist and teacher in theoretical elementary particle physics”, \textit{Quaderni di Storia della Fisica}, \textbf{1}, pp. 145–169.

\label{bon06} \item Bonetti, A. \&  Mazzoni, M. (2006). “The Arcetri School of Physics”, in Redondi, P. \textit{et al.} (eds), \textit{The Scientific Legacy of Beppo Occhialini}. Bologna : Società Italiana di Fisica; Berlin [etc.] : Springer, pp 3–34.

\label{bon07} \item Bonetti, A. \& Mazzoni, M. (2007). \textit{L’Università di Firenze nel centenario della nascita di Giuseppe Occhialini (1907-1993)}. Firenze: Firenze University Press.

\label{bon11} \item Bonolis, L. (2011). “Walther Bothe and Bruno Rossi: the birth and development of coincidence methods in cosmic-ray physics”, \textit{American Journal of Physics}, \textbf{79}(11), pp. 1133–1150.

\label{bon23} \item Bonolis, L., Buccella, F. \& Pancheri, G. (2023), “Raoul Gatto and Bruno Touschek's collaboration in the birth of electron-positron physics”, \textit{arXiv}, 2311.01293.

\label{car14} \item Cartacci, A. (2014). “The plates group of the Antonio Garbasso Institute of Florence (1953- 1983)”, \textit{Il Colle di Galileo}, \textbf{3}(1), pp. 7–14.

\label{cas23} \item Casalbuoni, R. (2023), “La statistica di Fermi – Fermi statistics”, \textit{Quaderni di Storia della  Fisica}, \textbf{28}(1), pp. 103-112. 

\label{cas18} \item Casalbuoni, R. \& Dominici, D. (2018). “The teacher of the gattini (kittens)”, \textit{Il Colle di Galileo}, \textbf{7}(2), pp. 47–69.

\label{cas21} \item Casalbuoni, R., Dominici, D. \& Mazzoni, M. (2021). \textit{Lo spirito di Arcetri A cento anni dalla nascita dell'Istituto di Fisica dell'Università di Firenze}. Firenze: Firenze University Press.

\label{cas22} \item Casalbuoni, R., Dominici, D. \& Mazzoni, M. (2022). “A brief history of Florentine physics from the 1920s to the end of the 1960s”, \textit{The European Physical Journal H}, \textbf{47}(1), p. 15.

\label{cas19} \item Casalbuoni, R., Dominici, D. \& Pelosi, G. (eds) (2021). \textit{Enrico Fermi a Firenze}. Firenze: Firenze University Press.

\label{cor00} \item Cordella, F. \& Sebastiani, F. (2000). “Sul percorso di Fermi verso la statistica quantistica”, \textit{Il Nuovo Saggiatore}, \textbf{16}, p. 11–22.

\label{goo01} \item Goodstein, J. (2001). A Conversation with Franco Rasetti. \textit{Phys. perspect.}, \textbf{3}, pp. 271–313.

\label{gue15} \item Guerra, F. \& Robotti, N. (2015). \textit{Enrico Fermi e il quaderno ritrovato 20 marzo 1934 La vera storia della scoperta della radioattività indotta da neutroni}. Bologna: SIF.

\label{lar20} \item La Rana, A. \& Rossi, P. (2020). “The blossoming of quantum mechanics in Italy: the roots, the context and the first spreading in Italian universities (1900-1947)”. \textit{The European Physical Journal H}, \textbf{45}, pp. 237–252.

\label{mai17} \item Maiani, L. \& Bonolis, L. (2017). “The Charm of Theoretical Physics (1958-1993), \textit{The European Physical Journal H}, \textbf{42}, pp. 611-661.

\label{man86} \item Mandò, M. (1986). “Notizie sugli studi di fisica (1859-1949)”, in \textit{Storia dell’Ateneo fiorentino}, volume 1. Firenze: Parretti Grafiche.

\label{ros87} \item Rossi, B., (1987). \textit{Momenti nella vita di uno scienziato}. Bologna: Zanichelli.

\label{tac17} \item Taccetti, N. (2017). “Physics with accelerators at Arcetri. A short chronicle dedicated to Tito Fazzini who was one of its leading protagonists”, \textit{Il Colle di Galileo}, \textbf{6}(1), pp. 19–38.

\end{bibliografia}

\end{Backmatter}

\end{document}